# Lattice-Boltzmann model for axisymmetric thermal flows

Q. Li, Y. L. He, G. H. Tang, and W. Q. Tao

National Key Laboratory of Multiphase Flow in Power Engineering,

School of Energy and Power Engineering, Xi'an Jiaotong University, Xi'an, Shaanxi 710049, China

In this brief report, a thermal lattice-Boltzmann (LB) model is presented for axisymmetric thermal flows in the incompressible limit. The model is based on the double-distribution-function LB method, which has attracted much attention since its emergence for its excellent numerical stability. Compared with the existing axisymmetric thermal LB models, the present model is simpler and retains the inherent features of the standard LB method. Numerical simulations are carried out for the thermally developing laminar flows in circular ducts and the natural convection in an annulus between two coaxial vertical cylinders. The Nusselt number obtained from the simulations agrees well with the analytical solutions and/or the results reported in previous studies.

In recent years, the lattice-Boltzmann (LB) method for simulating axisymmetric flows has attracted much attention [1-10]. In fact, the LB simulation of axisymmetric flows can be directly handled with a three-dimensional (3D) LB model. However, such a treatment does not take the advantage of the axisymmetric property of the flow: a 3D axisymmetric flow can be reduced to a quasi-2D problem. To make use of this feature, Halliday *et al*. [1] first studied the 2D LB method for axisymmetric flows in 2001. Some source terms containing density and velocity gradients were introduced into the microscopic evolution equation. However, this method fails to reproduce the correct hydrodynamic momentum equation due to some missing terms: the term $\rho u_i u_r / r$ is missing in the recovered momentum equation and some additional terms involving the first-order source term are missing in the second-order expansion of the microscopic evolution equation. These missing terms were noticed by Lee *et al*. [2] and Reis *et al*. [3, 4]. By adding these terms, Lee *et al*. developed a more accurate axisymmetric LB model. Reis *et al*. rederived Halliday *et al*.'s model and then presented a modified version. Zhou [5] recently proposed a simplified axisymmetric isothermal model, in which the source terms are simple, yet still contain a velocity gradient term which should be determined with a finite-difference scheme. Most recently, Guo *et al*. [6] developed a simple and consistent LB model for axisymmetric isothermal flows based on the continuous Boltzmann equation. The source terms in the model contain no gradients and are easier to implement.

There are also several attempts for constructing axisymmetric thermal LB models. The first attempt was made by Peng *et al*. [7] through the hybrid LB approach. In their model, the azimuthal velocity and the temperature field are solved by the second-order center difference scheme. Later, Huang *et al*. [8] found that, for flows with high Reynolds number and Rayleigh number, the convection terms in the Navier-Stokes equations become dominant and the second-order center difference scheme is unsuitable due to the enhanced numerical instability. Then they proposed an improved version of Peng *et al*.'s model. Recently, Chen *et al*. [9, 10] pointed out that, although Huang *et al*.'s hybrid LB model is more numerically stable than Peng *et al*.'s model, too many complicated source terms exist in their model and a great deal of lattice grids are still required for numerical stability. Noticing this problem, they devised a thermal LB model for axisymmetric thermal flows based on the vorticity-stream-function (VSF) equations [10]. The source terms are simplified by invoking the VSF formulation but still contain several gradient terms. Guo *et al.* recently argued that Chen *et al.*'s model will become very inefficient for unsteady flows because a Possion equation must be solved at every time step [6]. Meanwhile, the boundary condition is not easy to implement in the VSF-based numerical methods.

In the literature, Lallemand and Luo [11] have pointed out that the hybrid LB approach significantly deviates from the standard LB method, which means it loses some inherent features of the standard LB method, and it only provides a compromised solution. Alternatively, the double-distribution-function (DDF) LB approach [12-17], which utilizes two different distribution functions, one for the velocity field and the other for the temperature or energy field, has attracted much attention since its emergence for its excellent numerical stability as well as the retaining of inherent features of the standard LB method. The aim of this study is to develop a thermal LB model for simulating axisymmetric thermal flows based on the DDF LB approach. The velocity field of the incompressible axisymmetric thermal flows can be solved with isothermal axisymmetric LB models. In what follows we focus on discussing the microscopic evolution equation for solving the temperature field.

The macroscopic temperature equation of incompressible axisymmetric thermal flows in a cylindrical coordinate system can be written as

$$\partial_t T + u_i \partial_i T = \partial_i (\chi \partial_i T) + \chi \frac{1}{r} \partial_r T, \qquad (1)$$

where $T$ is the temperature; $i$ indicates the $r$ or $x$ component, here $r$ and $x$ are the coordinates in radial and axial directions, respectively; $u_i$ is the component of velocity in the $i$ direction; and $\chi$ is the thermal diffusivity. With the continuity equation $\partial_i u_i = -u_r/r$, we can

rewritten Eq. (1) as

$$\partial_t T + \partial_i (u_i T) = \partial_i (\chi \partial_i T) + \underline{\chi \frac{1}{r} \partial_r T - \frac{u_r T}{r}}. \tag{2}$$

The underlined terms arise from the cylindrical polar coordinate. In order to recover these terms, we introduce the following temperature evolution equation:

$$g_\alpha(\mathbf{r} + \mathbf{e}_\alpha \delta_t, t + \delta_t) - g_\alpha(\mathbf{r}, t) = -\frac{1}{2\tau_g}\left[g_\alpha(\mathbf{r} + \mathbf{e}_\alpha \delta_t, t + \delta_t) - g_\alpha^{eq}(\mathbf{r} + \mathbf{e}_\alpha \delta_t, t + \delta_t)\right]$$
$$-\frac{1}{2\tau_g}\left[g_\alpha(\mathbf{r},t) - g_\alpha^{eq}(\mathbf{r},t)\right] - \underline{\frac{e_{\alpha r}}{r}\delta_t \left[g_\alpha(\mathbf{r},t) - g_\alpha^{eq}(\mathbf{r},t)\right]}$$
$$+\frac{\delta_t}{2}\left[S_\alpha(\mathbf{r}+\mathbf{e}_\alpha\delta_t, t+\delta_t) + S_\alpha(\mathbf{r},t)\right], \tag{3}$$

where $g_\alpha$ is the temperature distribution function; $\tau_g$ is non-dimensional relaxation time for the temperature field; $S_\alpha$ is the source term; and the discrete velocities $\{\mathbf{e}_\alpha = (e_{\alpha x}, e_{\alpha r}): \alpha = 0,1,\ldots,8\}$ are specified by the standard D2Q9 lattice. The underlined term in Eq. (3) is used to recover the second term on the R.H.S of Eq. (2). Actually, we can prove that, when a similar treatment is combined with Zhou's isothermal axisymmetric model [5], the source terms of the rearranged model will contain no gradient terms.

$g_\alpha^{eq}$ is chosen as $g_\alpha^{eq} = T f_\alpha^{eq} = \rho T w_\alpha \left[1 + (\mathbf{e}_\alpha \cdot \mathbf{u})/c_s^2 + 0.5(\mathbf{e}_\alpha \cdot \mathbf{u})^2/c_s^4 - 0.5 u^2/c_s^2\right]$, where $c_s = c/\sqrt{3}$ ($c = \delta_x/\delta_t$ is the lattice speed) is the sound speed and the weights $w_\alpha$ are given by $w_0 = 4/9$, $w_{1-4} = 1/9$, and $w_{5-8} = 1/36$. It can be found that $g_\alpha^{eq}$ satisfies

$$\sum_\alpha g_\alpha^{eq} = \rho T, \quad \sum_\alpha e_{\alpha i} g_\alpha^{eq} = \rho T u_i, \tag{4}$$

$$\sum_\alpha e_{\alpha i} e_{\alpha j} g_\alpha^{eq} = \rho T u_i u_j + p T \delta_{ij}. \tag{5}$$

Through the second-order Taylor series expansion, the evolution equation (3) can be reduced to

$$\delta_t(\partial_t + \mathbf{e}_\alpha \cdot \nabla) g_\alpha + \frac{\delta_t^2}{2}(\partial_t + \mathbf{e}_\alpha \cdot \nabla)^2 g_\alpha = -\frac{1}{\tau_g}(g_\alpha - g_\alpha^{eq}) - \frac{\delta_t}{2\tau_g}(\partial_t + \mathbf{e}_\alpha \cdot \nabla)(g_\alpha - g_\alpha^{eq})$$
$$-\frac{e_{\alpha r}}{r}\delta_t(g_\alpha - g_\alpha^{eq}) + \delta_t S_\alpha + \frac{\delta_t^2}{2}(\partial_t + \mathbf{e}_\alpha \cdot \nabla) S_\alpha + O(\delta_t^3), \tag{6}$$

where $\nabla = (\partial_x, \partial_r)$ is the spatial gradient operator. By introducing the following expansions [18]

$$\partial_t = \partial_{t0} + \delta_t \partial_{t1}, \quad g_\alpha = g_\alpha^{(0)} + \delta_t g_\alpha^{(1)} + \delta_t^2 g_\alpha^{(2)}, \tag{7}$$

we can rewrite Eq. (6) in the consecutive orders of $\delta_t$ as

$$O(1): \quad g_\alpha^{(0)} = g_\alpha^{eq}, \tag{8}$$

$$O(\delta_t): \quad (\partial_{t0} + \mathbf{e}_\alpha \cdot \nabla) g_\alpha^{(0)} + \frac{1}{\tau_g} g_\alpha^{(1)} = S_\alpha, \tag{9}$$

$$O(\delta_t^2): \quad \partial_{t1} g_\alpha^{(0)} + (\partial_{t0} + \mathbf{e}_\alpha \cdot \nabla) g_\alpha^{(1)} + \frac{1}{2}(\partial_{t0} + \mathbf{e}_\alpha \cdot \nabla)^2 g_\alpha^{(0)} + \frac{1}{\tau_g} g_\alpha^{(2)} + \frac{1}{2\tau_g}(\partial_{t0} + \mathbf{e}_\alpha \cdot \nabla) g_\alpha^{(1)}$$

$$= -\frac{e_{\alpha r}}{r} g_\alpha^{(1)} + \frac{1}{2}(\partial_{t0} + \mathbf{e}_\alpha \cdot \nabla) S_\alpha. \tag{10}$$

Using Eq. (9), we can rewrite Eq. (10) as

$$\partial_{t1} g_\alpha^{(0)} + (\partial_{t0} + \mathbf{e}_\alpha \cdot \nabla) g_\alpha^{(1)} + \frac{1}{\tau_g} g_\alpha^{(2)} = -\frac{e_{\alpha r}}{r} g_\alpha^{(1)}. \tag{11}$$

Taking the summations of Eqs. (9) and (11), we can obtain, respectively

$$\partial_{t0}(\rho T) + \partial_j(\rho u_j T) = \sum_\alpha S_\alpha, \tag{12}$$

$$\partial_{t1}(\rho T) + \partial_i \left( \sum_\alpha e_{\alpha i} g_\alpha^{(1)} \right) = -\frac{1}{r} \sum_\alpha e_{\alpha r} g_\alpha^{(1)}. \tag{13}$$

To recover the target macroscopic temperature equation, $\sum_\alpha S_\alpha$ should be given by $\sum_\alpha S_\alpha = -\rho T u_r / r$. From Eq. (9), we have

$$\sum_\alpha e_{\alpha i} g_\alpha^{(1)} = -\tau_g \left( \partial_{t0} \sum_\alpha e_{\alpha i} g_\alpha^{(0)} + \partial_j \sum_\alpha e_{\alpha i} e_{\alpha j} g_\alpha^{(0)} \right) + \tau_g \sum_\alpha e_{\alpha i} S_\alpha. \tag{14}$$

From Eqs. (4), (5), and (8), we can obtain

$$\partial_{t0} \left( \sum_\alpha e_{\alpha i} g_\alpha^{(0)} \right) = u_i \partial_{t0}(\rho T) + \rho T \partial_{t0} u_i, \tag{15}$$

$$\partial_j \left( \sum_\alpha e_{\alpha i} e_{\alpha j} g_\alpha^{(0)} \right) = u_i \partial_j(\rho T u_j) + \rho T u_j \partial_j u_i + T \partial_i p + p \partial_i T, \tag{16}$$

where $\partial_{t0} u_i$ is evaluated as

$$\partial_{t0} u_i = \left[ \partial_{t0}(\rho u_i) - u_i \partial_{t0} \rho \right] / \rho = -u_j \partial_j u_i - (\partial_i p)/\rho. \tag{17}$$

According to Eqs. (15), (16), and (17), we can rewrite Eq. (14) as

$$\sum_\alpha e_{\alpha i} g_\alpha^{(1)} = -\tau_g \left[ u_i \partial_{t0}(\rho T) + u_i \partial_j(\rho T u_j) - \sum_\alpha e_{\alpha i} S_\alpha + p \partial_i T \right]. \tag{18}$$

If we carefully choose $\sum_\alpha e_{\alpha i} S_\alpha = -\rho T u_i u_r / r$, then Eq. (18) can be reduced to

$\sum_\alpha e_{\alpha i} g_\alpha^{(1)} = -\tau_g p \partial_i T$. Therefore in this study the source term $S_\alpha$ is chosen as

$$S_\alpha = -\frac{u_r}{r} g_\alpha^{eq}. \tag{19}$$

In the presence of a body force ($\boldsymbol{F} = \rho \boldsymbol{a}$), actually a forcing term $F_\alpha = \rho T w_\alpha (\boldsymbol{e}_\alpha \cdot \boldsymbol{a})/c_s^2$ should also be considered in Eq. (3). But this term seemingly can be neglected in most cases [12-14]. Substituting the equation $\sum_\alpha e_{\alpha i} g_\alpha^{(1)} = -\tau_g p \partial_i T$ into Eq. (13) and then combining Eq. (12) with Eq. (13) ($\partial_t = \partial_{t0} + \delta_t \partial_{t1}$), we can obtain the following macroscopic temperature equation:

$$\partial_t (\rho T) + \partial_i (\rho u_i T) = \partial_i (\rho \chi \partial_i T) + \rho \chi \frac{1}{r} \partial_r T - \frac{\rho u_r T}{r}, \tag{20}$$

where the thermal diffusivity $\chi$ is given by $\chi = \delta_t \tau_g c^2 / 3$. In the incompressible limit with $\rho \approx \rho_0$, Eq. (20) is just the target macroscopic temperature equation. To this end, we can simply modify $g_\alpha^{eq}$ as $g_\alpha^{eq} = \rho_0 f_\alpha^{eq} / \rho$. For small Mach-number flows, $g_\alpha^{eq}$ can be further simplified by neglecting the terms of $O(u^2)$ [14]. In this situation, $g_\alpha^{eq}$ based on a D2Q4 lattice with four directions $\boldsymbol{e}_1$, $\boldsymbol{e}_2$, $\boldsymbol{e}_3$, and $\boldsymbol{e}_4$ can also be used: $g_\alpha^{eq} = (T/4)\left[1 + 2(\boldsymbol{e}_\alpha \cdot \boldsymbol{u})/c^2\right]$ together with $\chi = \delta_t \tau_g c^2 / 2$.

To eliminate the implicitness of Eq. (3), following He et al. [13], a new distribution function $\overline{g}_\alpha = g_\alpha + 0.5(g_\alpha - g_\alpha^{eq})/\tau_g - 0.5\delta_t S_\alpha$ can be introduced. Through some standard algebra, the evolution equation for $\overline{g}_\alpha$ can be obtained

$$\overline{g}_\alpha(\boldsymbol{x} + \boldsymbol{e}_\alpha \delta_t, t + \delta_t) - \overline{g}_\alpha(\boldsymbol{x}, t) = -\omega_g \left[\overline{g}_\alpha(\boldsymbol{x}, t) - g_\alpha^{eq}(\boldsymbol{x}, t)\right] + (1 - 0.5\omega_g)\delta_t S_\alpha(\boldsymbol{x}, t), \tag{21}$$

where $\omega_g$ is given by $\omega_g = \left[1 + (e_{\alpha r} \tau_g \delta_t / r)\right] / (\tau_g + 0.5)$. The macroscopic temperature can be calculated from the new distribution function as $T = \sum_\alpha \overline{g}_\alpha / \left[\rho_0 (1 + 0.5\delta_t u_r / r)\right]$. The velocity field is solved by using isothermal axisymmetric LB models. Here Guo et al.'s isothermal axisymmetric LB model is adopted, which can be briefly summarized as follows [6]: the evolution equation for velocity field is

$$\tilde{f}_\alpha(\boldsymbol{x} + \boldsymbol{e}_\alpha \delta_t, t + \delta_t) - \tilde{f}_\alpha(\boldsymbol{x}, t) = -\omega_f \left[\tilde{f}_\alpha(\boldsymbol{x}, t) - \tilde{f}_\alpha^{eq}(\boldsymbol{x}, t)\right] + \delta_t (1 - 0.5\omega_f) G_\alpha(\boldsymbol{x}, t), \tag{22}$$

with $\omega_f = 1/(\tau_f + 0.5)$, $G_\alpha = (e_\alpha - u) \cdot \tilde{a} \tilde{f}_\alpha^{eq}/c_s^2$, $\tilde{a}_x = a_x$, $\tilde{a}_r = a_r + c_s^2[1 - 2\delta_t \tau_f u_r/r]/r$, and $\tilde{f}_\alpha^{eq} = r f_\alpha^{eq}$, where $\tau_f$ is the non-dimensional relaxation time for the velocity field; $a_x$ and $a_r$ are the components of the external force acceleration in the $x$ and $r$ directions, respectively. The macroscopic density and velocity are calculated by $\rho = \sum_\alpha \tilde{f}_\alpha / r$ and $u_i = r\left[\sum_\alpha e_{\alpha i} \tilde{f}_\alpha + 0.5\delta_t r \rho a_i + 0.5\delta_t \rho c_s^2 \delta_{ir}\right] / \left[\rho(r^2 + \tau_f \delta_t c_s^2 \delta_{ir})\right]$. The kinematic viscosity is given by $v = \delta_t \tau_f c^2 / 3$. Equations (21) and (22) together with the corresponding equilibrium distributions and the source terms constitute the present thermal DDF LB model for axisymmetric thermal flows.

Two numerical tests are considered to validate the proposed model. The first test is the thermally developing flow in a circular duct, which is a classical problem described in many heat transfer textbooks. A uniform temperature profile $T_{in} = 10$ and a thermally fully developed flow are respectively imposed at the inlet and outlet. Two different thermal boundary conditions (BC), the constant wall temperature BC (Type 1) and the constant wall heat flux BC (Type 2), are considered at the wall. In simulations, the relaxation time $\tau_f = 0.6$, the Prandtl number $Pr = v/\chi = 0.7$, and a grid size of $N_x \times N_r = 649 \times 82$ is adopted, corresponding to an aspect ratio $L/D = 648/81 = 8$, which is sufficient to describe a thermally developing flow. The periodic BC is applied in the axial direction for the velocity BCs with a body force $\rho a_x = \rho \times 10^{-4}$, while the non-equilibrium extrapolation BC is applied at the inlet and the wall for the thermal BCs. Meanwhile, the outflow is supposed to be fully developed and obeys the Neumann rule. The contours of the local Nusselt number, which is defined as $Nu_x = -D(\partial_r T)_w / (T_w - T_b)$, where $D$ is the diameter, $T_w = 1$, and $T_b = \int_0^{D/2} 2\pi r u T dr / \int_0^{D/2} 2\pi r u dr$ is the bulk temperature, are plotted in Fig. 1 along the axial direction. The Nusselt numbers are 3.674 and 4.376 respectively in the thermal fully developed region for the two different BCs. Compared with the corresponding analytical solutions [19], 3.66 and 4.36, the relative errors are 0.38% and 0.37%, respectively.

The second test is the natural convection in an annulus between two coaxial vertical cylinders [20, 21], which is a simplified representation of many practical problems and has been investigated both numerically and experimentally by a number of researchers. The problem is sketched in Fig. 2, where $g$ is the gravitation acceleration, $T_i$ and $T_o$ are the constant temperatures of the inner and outer cylinders, respectively, and $T_i > T_o$. The radius ratio $r_o/r_i$ and the aspect ratio $h/(r_o - r_i)$ are both set to be 2.0. The natural convection is characterized by the Prandtl number $Pr = \nu/\chi$ and the Rayleigh number $Ra = g\beta(T_i - T_o)(r_o - r_i)^3 Pr/\nu^2$, where $\beta$ is the thermal expansion coefficient. The buoyancy force is given by $\rho a_x = -\rho g(T - T_r)$, where $T_r = (T_i + T_o)/2$. Numerical simulations are carried out for $Ra = 10^4$ and $10^5$. A grid size of $N_r \times N_x = 101 \times 201$ is adopted. The streamlines and isotherms at the steady state are shown in Fig. 3 and 4, respectively. At a moderate $Ra$ number ($10^4$), a simple circulation is observed. When the $Ra$ number increases, the buoyancy force accelerates the circulation of fluid flow and the natural convection is significantly enhanced. As a result, the isothermals are greatly deformed. To quantify the results, in Table 1, the Nusselt numbers defined as $Nu_{i,o} = -(1/h) r_{i,o} \int_0^h (\partial_x T)_{i,o} / (T_i - T_o) dx$ are compared with the average Nusselt number ($\overline{Nu} = (Nu_i + Nu_o)/2$) reported in Refs. [20, 21]. Meanwhile, the average Nusselt numbers obtained from the D2Q4 lattice are 3.219 and 5.782 respectively for $Ra = 10^4$ and $10^5$, which are in good agreement with the results obtained from the D2Q9 lattice with $g_\alpha^{eq} = \rho_0 f_\alpha^{eq}/\rho$.

In summary, we have presented a thermal LB model for axisymmetric thermal flows based on the DDF LB approach. The source terms of the model contain no gradient terms. Compared with the existing axisymmetric thermal LB models, the present model is simpler and retains the inherent features of the standard LB method. Two numerical tests have been considered to validate the proposed model. Numerical results are compared with the analytical solutions and/or the results reported in previous studies. The comparisons show the capability and reliability of the model.

This work was supported by the Key Project of National Natural Science Foundation of China (No.50736005).

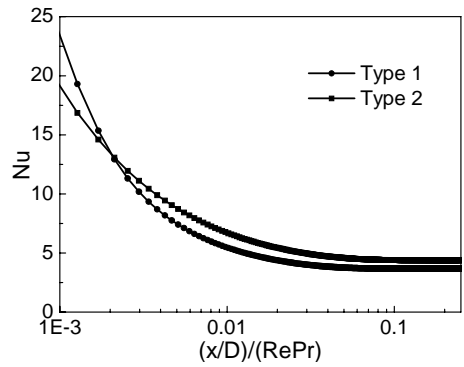

FIG. 1. Local Nusselt number distribution along the axial direction for the thermally developing flow.

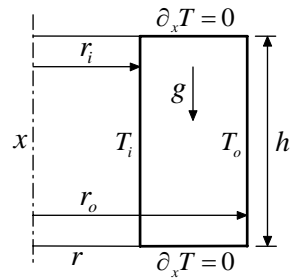

FIG. 2. Natural convection between coaxial vertical cylinders.

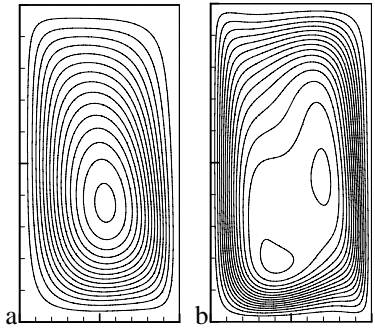

FIG. 3. Streamlines for $Ra = 10^4$ (a) and $10^5$ (b).

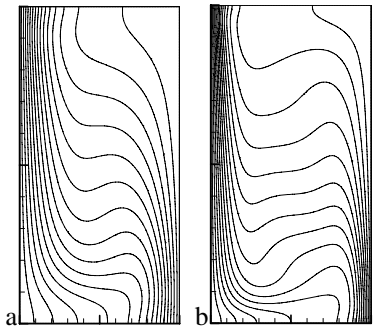

FIG. 4. Isotherms for $Ra = 10^4$ (a) and $10^5$ (b).

Table. 1. Comparison of the Nusselt number.

| $Ra$ | Ref. [20] | Ref. [21] | $Nu_i$ | $Nu_o$ |
|---|---|---|---|---|
| $10^4$ | 3.037 | 3.163 | 3.216 | 3.218 |
| $10^5$ | 5.760 | 5.882 | 5.782 | 5.787 |